\documentclass[twocolumn,showpacs,preprintnumbers,amsmath,amssymb]{revtex4}
\usepackage[english]{babel}

\usepackage{graphicx}

\newcommand{\beq}{\begin{equation}}
\newcommand{\eeq}{\end{equation}}
\newcommand{\bma}{\begin{math}}
\newcommand{\ema}{\end{math}}
\newcommand{\beqa}{\begin{eqnarray}}
\newcommand{\eeqa}{\end{eqnarray}}

\def\opone{\le\textbf{}\textbf{}avevmode\hbox{\small1\kern-3.8pt\normalsize1}}

\newcommand\vm[1] {V_1(z_{#1}) }
\newcommand{\av}[1]{\langle #1\rangle}

\newcommand\jas[3]{(z_{#1} - z_{#2})^{#3}}

\newcommand{\pref}[1]{(\ref{#1})}

\newcommand{\etal} {{\em et~al.} } 
\newcommand{\ie} {{\em i.e.} }
\newcommand{\eg} {{\em e.g.} }

\def\a{\alpha}
\def\b{\beta}
\def\g{\gamma}
\begin{document}

\title{ Hierarchy wave functions--from conformal correlators to Tao-Thouless states}

\author{E.J. Bergholtz$^1$}
\author{T.H. Hansson$^1$}
\author{M. Hermanns$^1$}
\author{A. Karlhede$^1$}
\author{S. Viefers$^2$}
\affiliation{$^1$Department of Physics, Stockholm University, AlbaNova University Center, SE-106 91 Stockholm,
Sweden \\
$^2$Department of Physics, University of Oslo, Box 1048 Blindern, NO-0316 Oslo, Norway}

\newcommand{\be}[1]{ \begin{eqnarray} \mbox{$\label{#1}$} }


      
\newcommand {\ee}{\end{eqnarray} }

\date{\today}

\begin{abstract} 
Laughlin's wave functions, describing the fractional quantum Hall effect at filling factors $\nu=1/(2k+1)$, can be obtained as correlation functions in 
conformal field theory, and recently this construction was extended to Jain's composite fermion wave functions at filling factors $\nu=n/(2kn+1)$. 
Here we generalize this latter construction and present ground state wave functions for all quantum  Hall 
hierarchy states that are obtained by successive condensation of quasielectrons (as opposed to quasiholes) in the original hierarchy construction. 
By considering these wave functions on a cylinder, we show that they approach the exact ground states, the Tao-Thouless states,  when the cylinder becomes thin. 
We also present wave functions for the multi-hole states, make the connection to Wen's general classification of abelian quantum Hall fluids, and discuss whether the fractional statistics of the quasiparticles can be analytically determined. 
Finally we discuss to what extent our wave functions can be described in the language of composite fermions.

\end{abstract}

\pacs{73.43.Cd, 71.10.Pm}

\maketitle

\section{Introduction}

Around 1990 it was noted that Laughlin's wave functions \cite{laughlin83} for the fractional quantum Hall (QH) effect take the form of 
correlation functions in conformal field theories (CFT) \cite{fubini, mooreread, cft,lutken}.  
It was conjectured that this is true for general QH-states, and that the 
same CFT that gives the ground state also describes the edge excitations \cite{wen,mooreread}.
Several other examples of wave functions that can be written as conformal correlators were given, and in particular the 
pfaffian  which has quasiparticles that obey non-abelian fractional statistics was proposed \cite{mooreread}.  This wave function 
is now believed to describe the observed gapped state at $\nu=5/2$\cite{5over2}. 
Recently it was shown that the composite fermion wave functions in the Jain sequence $\nu = n/(2kn+1)$\cite{jain89,jainbook}, 
$k,n=1,2,\ldots$ can be constructed from correlators in a CFT with $n$ bosonic fields \cite{hans}. 

The Laughlin state at $\nu_1=1/t_1$, $t_1= 3,5,\ldots$ has quasihole and quasielectron excitations with fractional 
charge $e^*=\pm e/t_1$.  According to the original hierarchy scheme \cite{haldane83,hierarchylaughlin,halperin84}, these 
quasiparticles may condense and form new fractional QH states at filling factors $\nu_2=1/(t_1\pm 1/t_2)$, $t_2=2,4,6,\ldots$ 
in much the same way as the electrons condense 
to form the Laughlin states. 
In the case of quasihole condensation we must also include the $t_1=1$ parent state corresponding to a filled Landau level. The quasiparticles in these
new QH states may then condense forming new states---a procedure that can be repeated ad infinitum, producing a unique  
QH state at each rational filling factor $\nu=p/q\le 1$, where $q$ is odd. Such a hierarchy state is characterized by the set
$\{ t_1, \alpha_2 t_2, \alpha_3 t_3, \dots \alpha_n t_n \}$, 
where the level of the hierarchy $n=1,2,\ldots$ is the number of condensates, $\alpha_i =\pm 1$ depending on whether condensate $i$
consists of quasiholes or quasielectrons.
Note that $t_1$ is odd, whereas $t_i=2,4,\ldots$, $i\ge 2$, are even. 

In section \ref{sec:gs}, we present candidate wave functions for all hierarchy states that are obtained by successive condensation of quasielectrons, as 
opposed to quasiholes, \ie for the states with $\alpha_i=-1$:  $\{ t_1, - t_2, \ldots ,-t_n \}$.
The wave functions are obtained using conformal field theory: a state at level $n$ is a correlation function in a theory with $n$ bosonic fields. 
If $\alpha_i t_i=-2$ for $i=2,\ldots, n$, we get the Jain sequence $\nu_n=n/((t_1-1)n+1)=n/(2kn+1), k=1,2,\ldots$, and the construction here reduces to the one presented in Ref. \onlinecite{hans} 
where it was shown that these wave functions are {\em identical} to Jain's composite fermion wave functions. The $t_i$'s determine the 
densities of the $n$ condensates that build up the state---the composite fermion state at level $n$ is the one where all but the first of these condensates 
have maximal density; when $t_i$ increases the density decreases. The wave functions presented here are thus obtained by a very natural generalization of the construction 
that gives the well-established ground state wave functions for the Jain fractions. Indeed, for general $\{ t_i \}$, the wave functions may alternatively be interpreted in terms of {\it fractional} quantum Hall states of  composite fermions.

It has been noted that the QH problem is exactly solvable at any rational filling factor in a certain limit, and it has been argued that 
this solution is adiabatically connected to the experimentally realized fractional QH state \cite{bkall,we07,bk3}. This limit can be obtained by considering 
the (polarized) electron gas on a cylinder or torus---this gives a natural mapping to a one-dimensional system---and letting the 
cylinder become thin. The electron-electron interaction then becomes 
purely electrostatic, and the ground state is a one-dimensional gapped crystal. Equivalently, one may consider 
this as the original infinite two-dimensional system with a modified electron-electron interaction. We call this limit the 
Tao-Thouless (TT) limit, and the ground states Tao-Thouless states, since the solution in this limit, for a Laughlin fraction, 
is the state proposed by Tao and Thouless as an explanation for the fractional QH effect \cite{tt}. In the TT-limit the hierarchy construction 
of QH states as successive condensates of quasiparticles is manifest. 

We have candidate bulk wave functions for hierarchy states that consist of condensates of quasielectrons only, but states involving condensations of quasiholes
also exist. In the TT-limit, there is a TT-ground state at each $\nu=p/q, \, q$ odd, that we believe corresponds to a potential bulk fractional QH state. In the 
occupation number representation used in the TT-limit, the particle-hole symmetry of the lowest Landau level is manifest. This is not the case for the wave functions, and 
we believe that it is for this technical  reason that we have not been able to obtain explicit wave functions for condensates involving quasiholes. 

In section \ref{sec:tt} we show that the hierarchy wave functions  $\{ t_1,  -t_2, \ldots, -t_n \}$ that we construct
using conformal techniques, reduce to the TT-states in the TT-limit, where they are constructed by $n$ successive condensations of (quasi)electrons.
In section \ref{sec:qp} we present the general quasihole wave functions and discuss their properties. Here we also explain how our wave functions fit into Wen's classification of abelian quantum Hall fluids. Finally, in 
 section \ref{sec:CF} we discuss the relationship between our proposed hierarchy wave functions and the composite fermion scheme.

The new ground state wave functions discussed here have been briefly reported in Ref. \onlinecite{we07}.  Several earlier approaches to hierarchy wave functions, in addition to Jain's composite fermions,  exist. In this context, we would like to refer to the construction of  composite fermion wave functions directly  in the lowest Landau level by Ginocchio and Haxton \cite{haxton} and the work by Wojs, Quinn, and collaborators, see Ref. \onlinecite{wojs} and references therein. Hierarchy wave functions in terms of quasiparticle coordinates were obtained by Moore and Read using conformal techniques \cite{mooreread}, whereas explicit wave functions were obtained by other methods by Greiter \cite{greiter94};  a conformal approach was used by Flohr and Osterloh \cite{flohr}.

\section{Hierarchy wave functions}
\label{sec:gs}
We here construct a unique ground state for each filling factor that is obtained, within the original hierarchy scheme,
by successive condensation of quasielectrons.  The state is obtained using conformal techniques generalizing the construction of the  Laughlin states \cite{mooreread} and the Jain states \cite{hans}. 

First we recall the conformal field theory approach to the Laughlin states \cite{mooreread}.   On the plane, with complex coordinates 
$z$, we introduce the vertex operator 
\beqa\label{vertexoperator1}
V_1(z) = :e^{i\gamma_1 \varphi_1(z)}: \ \ , 
\eeqa
where the normal ordering symbol : : will be suppressed in the following. $\varphi_1(z)$ is a free 
massless  holomorphic bosonic field, normalized so that the propagator becomes
\beqa\label{propagator}
\langle \varphi_1(z) \varphi_1 (w) \rangle =- {\rm ln} (z-w) \ \ .
\eeqa
This implies that vertex operators obey the relation
\beqa\label{ope}
e^{i\alpha \varphi_1(z)}e^{i\beta \varphi_1(w)}&=&e^{i\pi \alpha \beta} e^{i\beta \varphi_1(w)}e^{i\alpha \varphi_1(z)} \nonumber \\
&=&(z-w)^{\alpha \beta} e^{i\alpha \varphi_1(z)+i\beta \varphi_1(w)} \nonumber \\
&\sim& (z-w)^{\alpha \beta} e^{i(\alpha +\beta ) \varphi_1(w)} \ \ ,
\eeqa
where the last line is the operator product expansion valid as $z\rightarrow w$. Note that $e^{i\alpha \varphi_1(z)}$ and 
$e^{i\beta \varphi_1(w)}$ (anti)commute if $\alpha \beta$ is even (odd). 

The Laughlin wave function is obtained as a 
correlation function in the CFT of a string of radially ordered operators
\beqa\label{laughlin}
\av{ \vm 1 \vm 2 \dots  \vm N } 
=  \prod_{i<j} \jas i j {\gamma_1^2} e^{-\sum_i \frac { |z_i|^2} {4\ell^2}} \, ,
\eeqa
where $\av {\dots}$ is an expectation value in a vacuum state with an appropriately chosen background charge\cite{mooreread,hans}.
The crucial polynomial part follows directly from the operator product expansion in \pref{ope},  whereas the gaussian factor is 
obtained from the background charge. In the following we will suppress all gaussian factors. Equation \pref{laughlin} gives the Laughlin 
wave functions at $\nu=1/t_1$ provided
\beqa\label{gamma1}
\gamma_1=\sqrt {t_1}, \ \ t_1= 3,5,\ldots \ .
\eeqa
We note that these values are precisely the ones that make the operator $V_1 = e^{i\gamma_1 \varphi_1(z)}$  anticommute with itself; 
it is interpreted as an electron creation operator. 

To obtain wave functions for hierarchy states at level $n$ we define new operators recursively:  
\be{Vrecursion}
V_{\a+1}=\partial V_{\a} e^{-i{ \varphi_{\a}/ \gamma_\a}} e^{i{\gamma_{\a+1} \varphi_{\a+1}}}, 
\ee
for $\a=1,2,\ldots, n-1$, where  $\varphi_{\a+1}$ is a new bosonic field obeying \pref{propagator}. 
Formally, $V_{\a+1}$ are descendants of primary fields---the latter are exponentials of bosonic fields only, without derivatives.
It can be shown analytically  that, when going
from any given level in the hierarchy to the next, it is necessary to introduce this additional 
partial derivative in order to produce non-zero wave functions. We shall elaborate on this point in
section \ref{sec:CF}.
The vertex operators obey operator product expansions
\beqa\label{Vope}
V_\a(z)V_\a(w) &\sim'& (z-w)^{s_\a} \\ \nonumber
V_\a(z)V_\b(w) &\sim'& (z-w)^{s_{\a \b}}  \  ,
\eeqa
where $\sim'$ indicates that we have suppressed the derivatives. From \pref{Vrecursion} we find
\beqa\label{ks}
s_{\a+1}&=&s_\a+\gamma_\a^{-2}+\gamma_{\a+1}^2 - 2  \nonumber \\
s_{\a \b}&=&s_{\b \a}=s_\a-1 \ , \ \ \ \ {\rm for} \ \b >\a \ \ .
\eeqa
From \pref{gamma1} we see that $s_1=t_1= 3,5,\ldots$ is an {\it odd} positive integer. We require this to be true for all $s_\a$, for  the operators  
$V_\a$ to anticommute; from \pref{ks} we see that this implies that $\gamma_\a^{-2}+\gamma_{\a+1}^2$ is an {\it even} positive integer $t_{\a+1}$, thus
\beqa\label{gammarecursion}
\gamma_{\a+1}= \sqrt{t_{\a+1}-\gamma_\a^{-2}}\ , \ \  t_{\a+1}=2,4,6,\ldots \ .
\eeqa
The exponents $s_\a, s_{\a \b}$ in the operator product expansion \pref{Vope} then become
\beqa\label{k}
s_\a=s_{\a \b}+1=\sum_{\g=1}^\a t_\g -2(\a-1) \ , \ \ \b >\a \ \ .
\eeqa

The wave function is obtained as the conformal correlator 
\beqa\label{psi}
\Psi ={\cal  A}  \langle\prod _{\alpha=1} ^{n}\prod_{i_\alpha=1} ^{M_{\alpha}}V_\alpha (z_{i_\alpha}) \rangle \ , 
\eeqa
where $\cal A$ denotes antisymmetrization, and $M_{\alpha}$ is the number of particles in subset $\alpha$. 

It is straightforward to evaluate the correlators in \pref{psi}, and the explicit wave functions are in fact fully determined by the short distance behavior coded in the operator product expansion \pref{Vope}.\footnote{
This is not true on a topologically nontrivial manifold like the torus, where a careful evaluation of the correlators yields extra factors in addition to the periodized version of the wave functions presented here.}
The coordinates are divided into $n$ sets with $M_\a$ particles in each, and the exponents in the polynomials are 
$s_\a$ if both coordinates belong to set $\a$ and $s_{\a \b}$ if the two coordinates belong to the different sets $\a$ and $\b$; 
Equation (\ref{k}) gives $s_\a$ and $s_{\a \b}$ in terms of $t_\a$.  
At level $n$ of the hierarchy, the wave function is 
\be{psin}
\Psi &=&{\cal A} \{ (1-1)^{s_1} \partial_2 (2-2)^{s_2}\cdots  \partial_n^{n-1}(n-n)^{s_n}  \\
&\times&  (1-2)^{s_{12}} (1-3)^{s_{13}} \cdots ((n-1)-n)^{s_{n-1,n}} \} \ , \nonumber
\ee
where 
\beqa\label{notation}
\partial_\a^{\a-1} (\a-\a)^{s_\a} &\equiv&  \prod_{i_\alpha=1}^{M_\a} \partial_{z_{i_\alpha}}^{\a-1} 
\prod_{i<j\in M_\a}(z_i-z_j)^{s_\a}  \nonumber
\\
(\a-\b)^{s_{\a \b}} &\equiv&  \prod_{i_\alpha=1}^{M_\a} \prod_{i_\beta=1}^{M_\b} (z_{i_\alpha}-z_{i_\beta})^{s_{\a \b}} \ ,
\eeqa
and $z_{i_\a}$ numbers the $M_\a$ coordinates in set $\alpha$. (The derivatives in \pref{notation} act on all of $\Psi$ in \pref{psin}.)

From (\ref{psin}) and (\ref{notation}) we find that the highest power of a coordinate in subset $\alpha$ is 
\beqa\label{M1}
s_\a (M_\a-1)+\sum_{\b \neq \a}^n s_{\a \b}M_\b -(\a-1) \ \ .
\eeqa
The numbers of particles, $M_\a$, in the subsets are determined by requiring the highest power in each subset to be equal (up to terms 
of order one)
\beqa\label{M2}
s_\a M_\a+\sum_{\b \neq \a}^ns_{\a \b}M_\b= const. \ \ ,
\eeqa
where the constant is independent of $\a$. This corresponds to the different sets of particles having the same size.
Using \pref{k},  \pref{M2} implies
\beqa\label{Mrecursion}
M_\a=M_{\a+1}+(t_{\a+1}-2)\sum_{\b=\a+1}^nM_\b \ \ ,
\eeqa
for $\a=1,2,\dots n-1$, from which one easily obtains $M_\a$ in terms of $M_n$ for given $n$. 
The filling factor of $\Psi$ is 
\beqa\label{nu}
\nu_n=\frac 1 {t_1-\frac {1} {t_2-\frac {1} {  \frac {{\cdot_\cdot} }    {t_{n-1}-\frac {1} {t_n} }         }   }} \ \ ,
\eeqa
in accordance with the hierarchy construction\cite{haldane83,halperin84}. This is determined in the TT-limit below but can also be obtained by counting the number of particles 
in the wave function \pref{psi} and calculating the area it covers.

\section{Tao-Thouless limit} \label{sec:tt}

To obtain the TT-limit of a wave function we first translate it to the cylinder\cite{cyl,Haldane94} and then let the circumference of the cylinder go to zero;  for details see Ref. \onlinecite{bk3}.  This may equivalently be viewed as changing the hamiltonian while keeping the two-dimensional space infinite. 
The first step is achieved by the replacement
\beqa
z_i \rightarrow \beta_i \equiv e^{2\pi i z_i/L_1}
\eeqa
in the polynomial part of the wave function. (In addition, the gaussian 
factor is changed according to: $e^{-\sum_i|z_i|^2/{4\ell^2}} \rightarrow e^{-\sum_i |y_i|^2/{2\ell^2}}$. This factor is unaffected by 
the limiting procedure.) $z_i=x_i+i y_i$ are now complex coordinates 
on a cylinder with circumference $L_1$ in the $y$-direction. 
A basis of lowest Landau level states is given by
\beqa\label{cylstate}
\psi_k&=&\pi^{-1/4}L_1^{-1/2}    e^{2\pi i \frac{k x}{L_1}   } e^{-(y+k2\pi\ell^2/L_1)^2/{2\ell^2}} \nonumber \\ 
&\propto& e^{-2(\frac {k\pi\ell} {L_1})^2} \beta ^k \ , \ \  \ \ k=0,\pm1,\pm2,\dots
\eeqa
Since $\psi_k$ is centered at $y=-k2\pi\ell^2/L_1$, this maps the lowest Landau level onto a one-dimensional lattice model,
where the momentum in the $x$-direction, $2\pi k/L_1$, numbers the sites. 
The next step is to translate the many-particle wave function to the occupation number basis by expressing it in 
terms of the single-particle wave functions $\psi_k$ using \pref{cylstate}. For a generic term in the polynomial we have
(in the following we put $\ell = 1$)
\beqa
\prod _i \beta_i^{k_i } \propto e^{2(\frac {\pi} {L_1})^2\sum_i k_i^2}\prod _i \psi_{k_i}(z_i)\ \ ;
\eeqa
hence, when $L_1 \rightarrow 0$, the wave function approaches the occupation number state which maximizes 
$\sum_i k_i^2$. The momentum $K=\sum_i k_i$ is conserved---all components of the wave function in the occupation 
number basis have the same $K$.
Thus the task is to find the term in the polynomial \pref{psin} that has largest possible $\sum_i k_i^2$ 
for given $K=\sum_i k_i$. Since the state is fermionic, all $k_i$ will be different. In fact, all we have to do is to find 
one term  that maximizes $\sum_i k_i^2$  among the terms where all $k_i$ are different; the partners giving the antisymmetrization will be there automatically, and the terms with equal $k_i$'s cancel by antisymmetrization.
The term with the obtained set $\{k_i\}$ gives the state in the TT-limit: the electrons occupy the sites $\{k_i\}$ on the 
one-dimensional lattice, and the  wave function is the single Slater determinant ${\rm Det} [\psi_{k_i}(z_j)]$.  

The term that maximizes $\sum_i k_i^2$ for given $K=\sum_i k_i$ is obtained by first finding a coordinate with the largest $k_i$, 
then among the terms with this coordinate and this $k_i$ find a new coordinate with the next largest $k_j$ and so on. 
(Note that the coordinates in the different sets have different maximal $k$, and that we may assume that $k_i >0$, for all $i$,  since shifting all $k_i$'s by the same constant just amounts to a rigid translation of the state along the cylinder.)

Subtracting the common constant in \pref{M2} from \pref{M1}, we find that the 
highest power of a coordinate in subset $\a$ is
\beqa\label{highest}
\kappa_\a \equiv -s_\a-(\a-1) = -\sum_{\b=1}^\a t_\b+\a -1 \ \ \ . 
\eeqa
We now pick a coordinate with the highest $\kappa_\a$ and restrict to the terms that contain this coordinate to this power 
$ k_1 \equiv {\rm max} \, \kappa_\a$. We then determine the highest powers in the subsets $\a$ of the remaining coordinates 
in these terms  by inspecting the wave function in \pref{psin}. 
These powers are obtained from the ones defined in \pref{highest} by $\kappa_\a \rightarrow \kappa_\a+\delta \kappa_\a$, where 
\beqa\label{deltakappa}
\delta \kappa_\g&=&-s_\g   \nonumber \\
\delta \kappa_\b&=&-s_{\b \g} \  , \ \  \ \b \neq \g \ \ ,
\eeqa
where $\g$ is the subset that had the highest power. We again pick a coordinate with the highest $\kappa_\a$ and 
restrict to the terms that contain this coordinate to this power 
$ k_2\equiv {\rm max} \, \kappa_\a$.  Repeating this procedure we eventually find the set 
$\{k_1, k_2,\dots  k_N\}$ that gives the state in the TT-limit.

To reveal the general structure of the hierarchy it is convenient to introduce the differences
\beqa
\Delta_\a\equiv \kappa_1-\kappa_{\a} \ , \ \  \alpha =2,3\dots n \ \  .
\eeqa
From \pref{highest}, we find their initial values
\be{begin}
\Delta_\a^{(0)} = \sum_{\b=2}^\a t_\b-(\a-1) \ \ ,
\ee
which are positive and increasing $0< \Delta_\a <\Delta_{\a+1}$. From \pref{deltakappa} we find how $\Delta_\a$ changes.  
Letting $\g$ be the subset with the highest power, we find \beqa
\g=1: \ \ \ \ \ \ \delta \Delta_\a &=& s_{1\a}-s_1   \nonumber \\
\g \ge 2: \ \ \ \ \ \  \delta \Delta_\g&=&s_{\g}-s_{1\g} \nonumber \\
\delta \Delta_\a&=&s_{\a\g}-s_{1\g} \  , \ \  \ \a \neq \g \ \ ;
\eeqa
in terms of $t_\a$, this becomes
\beqa\label{dD}
\g=1: \ \ \ \ \ \  \delta \Delta_\a &=& -1  \nonumber \\
\g \ge 2: \ \ \ \ \ \  \delta \Delta_\a &=& \sum_{\b=2}^\a t_\b -2(\a -1) \ \ , \ \ \a <\g \nonumber \\
\delta \Delta_\g &=& \sum_{\b=2}^\g t_\b -2(\g -1) +1\ \ , \ \ \  \nonumber \\
\delta \Delta_\a &=& \delta \Delta_\g -1  \ \ , \ \ \ \ \ \ \ \  \ \ \a >\g \ \ .
\eeqa

We are now ready to determine the states in the TT-limit. At the first level of the hierarchy,  $n=1$, there is only one subset of coordinates in \pref{psin}
and the highest power changes by $\delta \kappa_1 = -t_1$ in each step, according to \pref{deltakappa}. Thus $\{k_i\}=\{0,-t_1,-2t_1,-3t_1,\dots \}$, 
where we have arbitrarily chosen $k_1=0$. In terms of the one-dimensional lattice this is a periodic system with one electron on every $t_1$:th site, 
{\it i.e.}, a crystal with unit cell 
\beqa\label{uc1}
{\bf C}^{(1)}=0_{t_1-1}1 \ , \ \ \ \ \nu_1= \frac 1 {t_1} \ \  .
\eeqa
We use a notation where 1(0) denotes that a site is occupied(empty) and $0_21=001$ etc. These 
are the states proposed by Tao and Thouless for the fractional quantum Hall effect\cite{tt}; we see that they are the TT-limits of the Laughlin states\cite{Haldane94}.

From the relations
\beqa
\delta \kappa_1&=&-t_1 \ \ , \ \ \  \g=1 \nonumber \\
\delta \kappa_1&=&-t_1+1 \ \ , \ \ \ \g \ge2  \ \ ,
\eeqa
which follow from
\pref{deltakappa}, the initial values $\Delta_\a^{(0)}$,  and \pref{dD}, we determine the sequence of $\{\kappa_1,\Delta_\a\}$'s at higher levels of the hierarchy.  These  give 
the largest $\kappa_\a$'s, \ie the $k_i$'s giving the state.  It turns out that independent of the initial value, the sequence of $\{\Delta_\a\}$'s is periodic, except  for possible edge effects that we ignore,
and that it contains the configuration $\{\Delta_\a\} = \{0,0,\dots 0,-1\}$\footnote{
Working out a few examples, this is fairly evident, and we have not bothered to construct a formal proof.}; 
 we will take this, rather than \pref{begin}, as the initial configuration when determining the sequence. In addition we set the initial largest $\kappa_\a$, which is $\kappa_1$, to zero.

At the second level, $n=2$, we find the results in Table \ref{table_1}.  For clarity we have included $\kappa_2=\kappa_1-\Delta_2$ explicitly.
The last column gives $\g$, the group which has the largest $\kappa_\a$.
The state is periodic since  $\{\Delta_2\}$ returns to its initial value at the last step. The highest $\kappa_\a$ at each step is shown in bold face, and these
give  $\{-k_i\}=\{(0),t_1,2t_1,3t_1,\dots (t_2-1)t_1,t_2t_1-1\}$ (Where $k_1=0$ is set in parenthesis 
since it belongs to the next unit cell. If two $\kappa_\a$ are equal 
we choose by convention the one with the smallest $\a$.)
This gives us the unit cell 
\beqa\label{uc2}
{\bf C}^{(2)}=\{0_{t_1-1}1\}_{t_2-1}0_{t_1-2}1 \ \ , \ \ \ \nu_2=\frac 1 {t_1-\frac {1} {t_2 }}  \ \ ,
\eeqa
where the filling factor is obtained by simply counting the number of ones and sites in ${\bf C}^{(2)}$.
Note that the unit cell at level one is repeated $t_2-1$ times in ${\bf C}^{(2)}$.

\begin{table}
\renewcommand{\arraystretch}{1.3}
\caption{Level two} \label{table_1} \centering
\begin{tabular}{cc|c|c}
	\hline
 $\kappa_1$    &   $\kappa_2$& $\Delta_2$   &$\gamma  , (\kappa_\g \, {\rm max})$  \\
	\hline
	\hline
 -1 & {\bf 0}&-1   &2 \\
\hline
 ${\bf -t_1}$ & $-t_1-t_2+2$ &$t_2-2$   &1\\
 ${\bf -2t_1}$ & $-2t_1-t_2+3$&$t_2-3$ &1 \\
. & .& . &.\\
. & .& .&. \\
. & .& . &.\\
${\bf  -(t_2-1)t_1}$ & $-(t_2-1)t_1$&0  & 1 \\
$ -t_2t_1$ & ${\bf  -t_2t_1+1} $&-1 &2\\
	\hline\end{tabular}
\end{table}

At the third level of the hierarchy, $n=3$, we obtain the sequences in Table \ref{table_2}. By comparing $\{\kappa_1,\Delta_2,\g\}$ in the two tables, 
we see that the structure at level two is first repeated $t_3-1$ times at level three, thus ${\bf C}^{(3)}={\bf C}^{(2)}_{t_3-1}{\bf a}$. Here ${\bf a}$ is 
obtained from the last part in Table \ref{table_2}; it differs from the ${\bf C}^{(2)}$-parts in that one ${\bf C}^{(1)}$ is missing, ${\bf a}={\bf C}^{(1)}_{t_2-2}0_{t_1-2}1$.
Thus the TT-states at level three are
\beqa
{\bf C}^{(3)}&=&\{  \{0_{t_1-1}1\}_{t_2-1}0_{t_1-2}1   \}_{t_3-1}    \{0_{t_1-1}1\}_{t_2-2}0_{t_1-2}1 \ \   \nonumber \\
\nu_3&=&\frac 1 {t_1-\frac {1} {t_2-\frac {1} { t_3      }   }} \ \ \ .
\eeqa
\begin{table}
\renewcommand{\arraystretch}{1.3}
\caption{Level three} \label{table_2} \centering
\begin{tabular}{c|cc|c}
	\hline
 $\kappa_1$    & $\Delta_2$ &  $ \Delta_3$ & $\gamma  , (\kappa_\g \, {\rm max})$   \\
	\hline
	\hline
-1 & 0 & -1&3 \\
\hline
 $-t_1$ & $t_2-2$&$t_2+t_3-4$ &1 \\
  .      &     .     & . &  . \\
 $-t_2t_1$      &  -1        &  $t_3-3$  & 2   \\
\hline
 $-t_2t_1+1-t_1$ & $t_2-2$         &    $t_2+t_3-5$        &1 \\
.      &     .     & . & .  \\
$-2t_2t_1+1$  & -1       & $t_3-4$       &   2\\
\hline
.       &   .       & . &  . \\
\hline
$(t_3-1)(-t_2t_1+1)-t_1$  &  $t_2-2$         &    $t_2-2$        &1 \\
.        &      .    & . &  . \\
$-t_3t_2t_1+t_3-1$ &-1    &     -1      &  2\\
\hline
$-t_3t_2t_1+t_3-t_1$ &$t_2-2$     & $t_2-3$        & 1  \\
.        &      .    & . & .  \\

$-t_3t_2t_1+t_3+t_1(1-t_2)$ &0      &    -1     &  3  \\
	\hline\end{tabular}
\end{table}

By pondering the relation \pref{dD} one realizes that this structure 
extends to general level $n$, and that the unit cells for the states in the TT-limit obey the relation
\beqa\label{Crel}
{\bf C}^{(n)}={\bf C}^{(n-1)}_{t_n-1}\overline{{\bf C}^{(n-2)}} \ \ ,  \ \ {\bf C}^{(0)}\equiv 0 \ , \  \  n=1,2,\dots \ .
\eeqa
Here ${\bf C}^{(n-1)}_{t}$ indicates that ${\bf C}^{(n-1)}$ is repeated $t$ times and $\overline {{\bf C}^{(n-2)}}$ is 
the complement of ${\bf C}^{(n-2)}$ in the unit cell ${\bf C}^{(n-1)}$, \ie 
$ {\bf C}^{(n-2)}\overline {{\bf C}^{(n-2)}}={\bf C}^{(n-1)}$. 
These unit cells are the ground states in the TT-limit\cite{bk3}, and the filling factors are the ones in \pref{nu}. Moreover, 
$\overline {{\bf C}^{(n-2)}}$ are the fractionally charged quasielectrons in the ground state with unit cell ${\bf C}^{(n-1)}$ and 
hence the state with unit cell ${\bf C}^{(n)}$ is, according to \pref{Crel}, a condensate of  the quasielectrons in the state with the 
unit cell  ${\bf C}^{(n-1)}$. Thus the original hierarchy construction is  manifest in the TT-limit.

\newcommand{\km} {{\bf K} }
\newcommand{\tv} {{\bf t} }
\newcommand{\tvt} {{\bf t^T} }
\newcommand{\cv} {{\bf c} }
\newcommand{\cvt} {{\bf c^T}} 
\newcommand{\qm} {{\bf Q} }
\newcommand{\lm} {{\bf L} }
\newcommand{\lmn} {{\bf L}_s }
\newcommand{\qv}[1] {{\bf Q}^{(#1)} } 
\newcommand{\lv}[1] {{\bf l}^{(#1)} } 
\newcommand{\lvt}[1] {{\bf l}^{(#1)}^T } 
\newcommand{\phiv} {{\bf\varphi} }

\section{Quasiholes and quasielectrons} \label{sec:qp}
In Ref. \onlinecite{hans} it was shown how to construct quasihole wave functions for the Jain series, and how to relate these to Wen's general classification\cite{wen} of abelian quantum Hall fluids. An abelian fluid (on a flat manifold) at level $n$  is specified by an $n\times n$ matrix $\bf K$, an $n$-dimensional charge vector $\bf t$, and $n$ distinct $n$-dimensional vectors $\lv \alpha$. Here we first generalize the quasihole construction to all the hierarchy states discussed above and give an explicit expression for the general multi-hole wave function. We then briefly discuss the quasielectron wave functions, but without giving explicit formulae. Next we make the connection to Wen's classification, and give explicit formulae for the quantities $\km$, $\tv$  and $\bf l$.
We end this section with a critical  discussion of the status of fractional statistics in the hierarchy states.

\subsection{Quasihole wave functions}
 First we express the $n$ electron operators \pref{Vrecursion}, at level $n$, in the form 
 \be{altop} 
 V_\alpha = \partial^{\alpha -1} e^{i \qv \alpha \cdot \phiv} \ \ ,
 \ee
which  is suitable for actually evaluating the correlators. Here $\phiv = (\varphi_1, \varphi_2, \dots \varphi_n)$, and $\bf Q^{(\alpha)}$ has components
$ Q^{(\alpha)}_\beta=q^{(\alpha)}_ \beta/R_\beta$, where $q^{(\alpha)}_ \beta$ are integers and $R_\beta$ is the compactification radius of the field $\varphi_\beta$.  
The $U(1)$ charge associated with the current 
$J(z) =i\sum_\alpha c_\alpha \partial \varphi_\alpha (z )=i \cv\cdot\partial\phiv$, where  the components of the "charge vector" $\bf c$ are
$c_\alpha = 1/R_\alpha$, measures the "depletion" of the quantum Hall fluid, and is thus closely connected to the electric charge. 
The condition  that the charges of the electron fields all equal one  (we shall give charge in units of the electron charge $-e$) leads to
the sum rule 
\be{sumrule}
\sum_{\beta=1}^n Q^{(\alpha)}_ \beta/R_\beta = 1  \ \ .
\ee
The explicit expressions for the integers $q^{(\alpha)}_ \beta$, which will not be needed here, can be found in Ref. \onlinecite{torus}. 
We also introduce the corresponding
 $n$ hole operators,
 \be{altop} 
 H_\alpha =  e^{i \lv \alpha \cdot \phiv} \ ,
 \ee 
where the vectors $ \lv {\alpha}$, $\alpha = 1,\dots n$, are reciprocal to the vectors $\qv \alpha$,
\be{rec}
\lv \alpha \cdot \qv \beta = \delta_{\alpha \beta}\ \ .
\ee

The wave function for a state with $m$ holes at positions $\{ \eta_i \}$ is given by the correlator
\be{mhole}
\Psi_h(\{\eta_i\}) ={\cal  A}  \langle    \prod_{i=1}^m H_{\beta_i} (\eta_i)   \prod _{\alpha=1} ^{n}\prod_{i_\alpha=1} ^{M_{\alpha}}V_\alpha (z_{i_\alpha}) \rangle \  \ , 
\ee
where antisymmetrization is performed over the electron coordinates $z_i $ only.
Evaluating the correlators gives
\be{mhexp}
\Psi_h &=&{\cal A} \{ (1-1)^{s_1} \partial_2 (2-2)^{s_2}\dots  \partial_n^{n-1}(n-n)^{s_n} \nonumber \\
&\times&  (1-2)^{s_{12}} (1-3)^{s_{13}} \dots ((n-1)-n)^{s_{n-1,n}}\nonumber  \\
&\times& f(\{\eta_i\} , \{{z_i}_\alpha\}     )  \} \ \ ,
\ee
where $s_{\alpha\beta} = \qv{\alpha} \cdot \qv {\beta} $, $s_\alpha \equiv s_{\alpha \alpha}$, and
\be{exprel}
 f(\{\eta_i\} , \{{z_i}_\alpha\}     ) &=& \prod_{\alpha = 1}^n \prod_{i_\alpha=1}^{M_\a} \prod_{i=1}^{m} (z_{i_\alpha}-\eta_i)^{\lv {\beta_i} \cdot \qv \alpha  } \nonumber  \\
 &\times& \prod_{i<j}^{n}  (\eta_i - \eta_j)^{ \lv{\beta_i} \cdot \lv {\beta_j} } \ \ .
 \ee

As stressed by Moore and Read, and discussed in some detail in Ref. \onlinecite{hans}, there is no local operator that will create quasielectrons. The naive guess, $H^{-1}_{\alpha}$, will give correlators which are not analytic in the {\em electron} coordinates, and thus not acceptable lowest Landau level wave functions. Nevertheless, one can  construct wave functions corresponding to many quasielectrons in specified angular momentum states. By forming coherent superpositions of such states one can then obtain localized multi-quasielectron wave functions. The explicit formulae are lengthy and not very illuminating, and we refer to Ref. \onlinecite{hans} for some explicit examples. Although there is no local quasielectron operator, it is possible to construct a {\em quasi-local} operator,  $P_{\alpha}(\eta)$,  with the same charge and conformal dimension as  $H^{-1}_{\alpha}(\eta)$, which directly creates the localized quasieletron wave functions\cite{qel}.

\subsection{Topological classification}

It is convenient to define the two matrices $\qm$ and $\lm$ with components 
\be{matdef}
{Q}_{\alpha \beta} = {Q}^{(\beta)}_ \alpha \ \ \ \mathrm {and} \ \ \  {L}_{\alpha \beta} = {l}^{(\beta)}_ \alpha \ .
\ee
The condition \pref{sumrule} then reads 
\be{elch}
\cv ^T \qm =  (1,1, \dots  1) \  ,
\ee
while the defining condition \pref{rec} takes the form
\be{rec2}
\lm^T\qm = \bf 1 \ .
\ee

The charges of the quasiparticles ${\bf q} = (q_1, q_2,\dots q_n)$ and the matrix ${\bf \Theta}$ containing the (mutual) statistical angles $\theta_{\alpha\beta}$ between two particles in the groups 
$\alpha$ and $\beta$ respectively, are given by the relations,
\be{qpch}
{\bf q} &=& -\cv^T   \lm   \\
{\bf \Theta } &=&\pi \lm^T \lm \label{theta} \  .
\ee
Also, as explained in Ref. \onlinecite{hans}, the filling fraction can be obtained from the background charges needed to make the correlators non-zero, and this yields the relation
\be{ffac}
\nu = \cv \cdot  \cv \ . 
\ee
The filling fraction $\nu$, together with the quasiparticle charges $\bf q$, and the statistical angles $\bf\Theta$, gives a complete macroscopic description of a quantum Hall fluid on an infinite plane.
Using \pref{elch} and \pref{rec2} the relations  \pref{ffac}, \pref{qpch} and  \pref{theta} can be expressed in the more familiar form given by Wen\cite{wen}: 
\be{finrel}
\nu &=&  \tv^T \km^{-1} \tv \nonumber \\
{\bf q} &=&  - \tv^T \km^{-1} \lmn \nonumber \\
{\bf \Theta} &=&   \pi  \lmn^T \km^{-1} \lmn  \ \ ,
\ee
if one defines 
\be{tlk}
\km &=& \qm^T\qm \nonumber \\
{\bf t}^T&=&(1,1, \dots 1) \nonumber \\
{\bf L}_s &=&\bf 1 \ \  .
\ee

The relations \pref{finrel} are precisely those introduced by Wen. 
Note that the columns of the unit vector ${\bf L}_s$, which we denote as ${\bf l}_s ^{(\alpha)}$, 
are orthogonal unit vectors. They are Wen's $\bf l$-vectors characterizing the $n$
fundamental quasiholes; a generic (composite) quasiparticle can be expressed as a linear
superposition of these.
The values of the vectors $\tv$ and ${\bf l}_s^{(\alpha)}$ correspond 
to his "symmetric basis". Also note that the entries in the $\km$-matrix only depend on the integer scalar products $\qv\alpha\cdot\qv\beta$, and thus on the powers of the Jastrow factors in the ground state wave function, but not on the specific choice of the vectors $\qv\alpha$. 

A Laughlin hole, which is created  by  the insertion of a thin unit flux tube, amounts to a unit vortex in all the $n$ condensates, and is given by the operator $H_L = \prod_{\alpha=1}^n H_\alpha$,
{\it i.e.}, the $\bf l$-vector $(1,1, ..., 1)$.
The expected values for statistics and charge, $\theta/\pi = -q = \nu$, follow from  \pref{finrel}. 

The quasiparticle states can also be constructed in the TT-limit \cite{bk3} and we suggest that a connection can be made to Wen's classification.

\subsection{Fractional statistics of the quasiparticles}
Above we tacitly assumed that the fractional statistics of the quasiholes could be read from the factors 
$(\eta_i - \eta_j)^{\theta_{ij}/\pi}$, which gives a  phase $2\theta_{ij}$ to the  wave function when one of the quasiparticles encircles the other.  This, however, is not the full story when considering a real physical process. First, the motion must be slow enough to be adiabatic, but more importantly, one might also pick up a Berry phase. It is only the sum of the Berry phase and the monodromy $2\theta_{ij}$ that has a physical significance and can be interpreted in terms of fractional statistics. In particular, if a wave function is normalized as to be  single-valued in the quasihole coordinates (it must always be single-valued in the electron coordinates), then the statistical phase equals the Berry phase. 

In the case of the Laughlin states, the Berry phase can be evaluated using several methods. In the original calculation by Arovas \etal it was directly related to the charge deficit \cite{arovas}, while Laughlin used the plasma analogy to evaluate the normalization of the two-quasihole wave function from which the Berry phase can be extracted\cite{laughlinstat,kjons}. There are also general arguments by Kivelson and Rocek \cite{stevemart} that relate fractional charge and fractional statistics for quasiholes where the electron density goes to zero in the core.
Furthermore, Su has argued that the statistics properties of the elementary quasiparticles follow from those of the Laughlin holes\cite{su86} if one assumes that the fractional statistics phase $\theta_C$ of a composite $C = A + B$  is given by $\theta_C = \theta_A + \theta_B + 2\theta_{AB}$, where $\theta_{AB}$ is the mutual statistics phase \cite{thwu}. 

That these very general arguments can be fallacious is demonstrated by the  $\nu = 1/2$ Moore-Read pfaffian state. While the Laughlin hole with charge 1/2 has statistics $\pi /2$, the elementary holes carry charge 1/4 and obey non-abelian statistics, which however is manifested only when four or more holes are present. The cause of this is a degeneracy of the electronic wave function for {\em fixed} positions of the quasiholes. This rather subtle effect is not at all obvious from the form of the pfaffian ground state wave function, but is readily seen in the structure of the quasihole operators in the conformal field theory framework. 

This last remark emphasizes a fact stressed  by Nayak and Wilczek \cite{naywil}: in all examples we know of, the expected fractional statistics is coded in the pertinent hole operators, \ie the statistical phases---abelian or non-abelian---are simply the monodromies of the conformal blocks in question; the Berry phase is simply zero. They also proposed that this is true in general, and they gave some arguments to support this hypothesis (see also Ref. \onlinecite{naygur}). From this perspective, the expression \pref{mhole} is interesting since the monodromies can be explicitly calculated (they are coded in the matrix $\bf\Theta$) while we know of no analytic method to calculate the Berry phases. (The reason for this is that the monodromies are identical for all terms in the sum implied by the antisymmetrization, while the Berry phase involves cross terms that are difficult to handle.) However, assuming that the Berry phases are zero for the wave functions \pref{mhexp}, which is supported by numerical calculations at $\nu = 2/5$ as well as the general arguments referred  to above, the statistics of the elementary holes in these hierarchy states is given by \pref{finrel}.

\section{Composite fermions}
\label{sec:CF}
To illuminate the relationship between the wave functions \pref{psin}, and those constructed using composite fermions, we first consider the  $n=2$ case,
\be{wavefns}
\Psi &=& {\cal A} \{  (1-1)^{t_1} \partial_2 (2-2)^{t_1 + t_2 -2} (1-2)^{t_1 - 1} \} \, ,
\ee
with groups 2 and 1 containing $M_2$ and $M_1 = M_2(t_2 - 1)$ particles, respectively. 
The derivatives act on everything to their right.
For $t_2=2$, $\Psi$ are Jain's wave function at $\nu = 2/(2t_1-1)$, 
while $t_1=3,t_2=4$ gives our proposal for the observed state at $\nu = 4/11$. 

For general $n$, if  $t_\a=2$ for $\a=2,\ldots, n$, it follows from \pref{nu} that $\nu_n=n/((t_1-1)n+1)$, 
and the wave functions 
\pref{wavefns} are  Jain's composite fermion wave functions as explained in Ref. \onlinecite{hans}. 
The $t_i$'s determine the densities of the $n$ condensates that build 
up the state---the "principal" composite fermion states are the ones where all but the first of these condensates have maximal density; in the CF language, they correspond to integer quantum Hall states of composite fermions.

More generally, one can interpret all the hierarchy states \pref{wavefns} in terms of
fractional quantum Hall states of composite fermions \cite{chang04}. This is most easily illustrated
by the case of $\nu=4/11$ obtained by taking  $t_1=3,t_2=4$:
\beqa\label{4/11}
\Psi_{4/11} &=& {\cal A} \{ (1-1)^{3} \partial_2 (2-2)^{5} (1-2)^{2} \} \nonumber \\
&=& {\cal A} \{ (1-1)^{1} \partial_2 (2-2)^{3} \} \prod_{i<j}(z_i-z_j)^2, 
\eeqa
where the squared Jastrow factor includes all pairs of particles. The first group of coordinates corresponds to the usual Slater determinant of the filled lowest CF Landau level, while the second describes the 1/3-filled second CF Landau level---it may be written as a sum of Slater determinants of the various allowed distributions of $M_2$ composite fermions in the $3M_2$ states of the second CF Landau level. 
In other words, the 4/11 state can be viewed as a $\nu^* = 1 + 1/3$ state of composite fermions, $\nu^{-1}=\nu^{*-1}+2=11/4$. This was first 
pointed out by Chang and Jain, who proposed a wave function which is similar to 
the one given here \cite{chang04}. 

Note that the derivatives $\partial_2$ which are necessary in order for this wave function to be
non-zero under antisymmetrization over the two groups \cite{hans}, are the lowest Landau level projection
of the anti-holomorphic coordinates $\bar z_2$ characteristic of the second Landau level.

Similar reasoning can be applied at higher levels in the hierarchy, where a
less trivial example is the $\nu=12/29$  state at level $n=4$, obtained from the set
$t_{\alpha} = 3,2,4,2$, giving $s_1 = s_2 = 3;~ s_3 = s_4 = 5; ~ s_{1\beta} =s_{2\beta} =2; ~ s_{34}=4 $.
Pulling out a full, squared Jastrow factor and rearranging derivatives, this wave function may be rewritten in the following, suggestive way: 
\beqa\label{12/29}
\Psi_{12/29} &=& {\cal A} \{ (1-1)^{1} \partial_2 (2-2)^{1} \partial_3^2 \partial_4^2  \\
 &\times &\left[   (3-3)^3 \partial_4 (4-4)^3 (3-4)^2  \right]   \} \prod_{i<j}(z_i-z_j)^2. \nonumber 
\eeqa
The CF interpretation of this state goes as follows: Groups 1 and 2 fill the two lowest CF Landau levels, while the 
part inside the square brackets has the form of a 2/5 CF state, consisting of the
groups 3 and 4, {\it cf} \pref{wavefns}. The  double derivatives can be thought of as  "lifting" this state to the third
CF Landau level. So the complete state can be interpreted
as the $\nu^* = 1 + 1 + 2/5$ CF state, $\nu^{-1}=\nu^{*-1}+2=29/12$. Again, it can be seen that the
wave function would be zero if $V_4$ did not have an additional derivative as compared to
$V_3$, since the expression in square brackets would be zero under antisymmetrization
of groups 3 and 4 in the absense of $\partial_4$. (Note that both the operator $\partial_3^2 \partial_4^2 $ and
the overall Jastrow factor is symmetric under exchange of 3 and 4.)

It is fairly straightforward to convince oneself that similar arguments, including
the necessity of an additional derivative at each level, carries over to the
entire hierarchy:
Each time a $t_{\gamma} > 2$ ($\gamma \ge 2$)  enters the sequence $\{ t_{\alpha} \}$, the
subsequent groups ($\beta \geq \gamma$) combine to form a fractional state of composite fermions
in the topmost CF Landau level, while the groups corresponding to $\beta < \gamma$ 
form filled CF Landau levels. For example, if only one of the $t$'s differs from 2,
$t_{\alpha} = 2;~\alpha \neq \gamma$, $t_{\gamma}=4$, the resulting state will correspond
to an integer (depending on the position of $t_{\gamma}$) number of filled CF Landau levels,
plus one of the principal Jain fillings (1/3, 2/5, 3/7...) in the topmost level; with $t_{\gamma}=6$
instead, the filling of the topmost level is one of the $k=2$ Jain fractions (1/5, 2/9, ...). 
If more than one of the $t_i$'s differs from 2, the filling of the topmost CF Landau level is in itself "hierarchical" ({\it e.g.},
the sequence $t_{\alpha} = 3,2,4,4$ corresponds to $\nu^* = 1+1+ 4/11$). 

Taking the picture of weakly interacting composite fermions literally, it is natural that only the topmost CF Landau level can be partially filled since the levels are split by an effective cyclotron gap. Needless to say, our wave functions have no notion of such a cyclotron gap since they are constructed directly in the lowest Landau level. However, it should be noted that an effective Landau level structure, with effective cyclotron gaps, emerges in the TT-limit. States corresponding to several partially filled CF Landau levels, as \eg the $n=2$ state at $\nu = 2/7$ with $s_1 =s_2 = 5$ and $s_{12} =2$, are not included in the hierarchy sequence \pref{wavefns}, and might in fact be non-abelian. 

Note that except for the Jain sequence, there is no direct connection between the hierarchy level $n$ and the number of occupied CF Landau levels, since the uppermost level can support an arbitrarily complicated fractional state. 
This also implies that as one moves up in the hierarchy, the CF interpretation of the states becomes increasingly contrived, and the appealing simplicity of the composite fermion description of the Jain sequence is all but lost.
 
In the above analogies, we were somewhat cavalier about exactly how the derivatives were acting, since this is anyhow not well defined within the CF scheme. (The wave functions for the positive Jain series discussed in this paper correspond to the originally proposed CF projection scheme.) A more interesting question is whether the derivatives in \pref{Vrecursion} are all needed, especially since the topologically  relevant quantities are all coded in the powers of the Jastrow factors. Here we only offered a partial answer: Within the hierarchy construction implied by \pref{Vrecursion}, the derivative can not be omitted as argued above. Although we can not fully exclude the possibility of {\em deleting} derivatives from  $V_n$ when constructing $V_{n+1}$, we find this possibility unlikely. The reason is that  $V_{n+1}$ is closely related to the quasielectron operator which can be shown to be quasi-local, thus excluding long-range operators such as inverse derivatives\cite{qel}.

The original hierarchy construction is manifest in the Tao-Thouless limit; however, a connection to composite fermions can be made also in this limit. 
The TT-states at filling factors 1/3, 2/5 and 4/11 are $0_21, \, 0_2101, \, (0_21)_301$ respectively. $0_21$ is the filled lowest effective (CF) Landau level.  
Quasielectrons 01 in a 1/3 ground state with $N$ cells $(0_21)_N$ can be inserted in $N$ equivalent places, which may be interpreted as an effective Landau level 
with $N$ states. Filling all these gives the state with unit cell $0_2101$, {\it i.e.}, the 2/5 state, which thus can be interpreted as consisting of two filled effective Landau levels. If instead one
third of the states in the second effective Landau level is filled, then the state with unit cell $(0_2 1)_3 01$, {\it i.e.}, the 4/11 ground state, is obtained. This interpretation can be extended to other filling factors but becomes less natural at higher levels in the hierarchy.\footnote{For example,
the TT-state at 12/29 is $(0_2101)_301(0_2101)_201$. As discussed in the text, two filled effective Landau levels,  $\nu^* = 1 + 1$, corresponds to $0_2101\equiv a$; letting 
$b=01$, we have $(0_2101)_301(0_2101)_201=a_3b a_2b$. Thus $b$ has been inserted in two out of five equivalent positions and hence one may interpret the 12/29 state 
as a $\nu^* = 1 + 1 + 2/5$ CF state. To minimize the energy, the $b$'s are as far separated as possible in the $a-$state. }

\section{Summary and outlook}
In this paper we constructed  wave functions for the quantum Hall hierarchy ground states that are built by successive condensations of quasielectrons, and explicitly showed that, when put on a cylinder,  they evolve into the Tao-Thouless ground states when the cylinder becomes thin. We also constructed the general quasihole states and showed, using standard assumptions about Berry phases, that they have all expected topological properties and fit into Wen's general classification of QH fluids. Finally we pointed out that the successful composite fermion approach to QH physics fits naturally in our scheme. 

There are many questions that have to be answered before we can claim to have a comprehensive understanding of the QH hierarchy states. First, our approach is so far limited to condensation of quasielectrons. It is not clear how to incorporate the condensation of quasiholes, but it is likely to involve products of holomorphic and non-holomorphic conformal blocks. These objects have nice topological properties, but it is less clear how to extract lowest Landau level wave functions. 

Another interesting open question concerns the possibility of having non-abelian states with wave functions very similar to \pref{psin} and we already mentioned the  one at $\nu = 2/7$.  

Finally, it is of course crucial to establish the connection to experimentally observed states which are believed to be abelian but do not belong to the main Jain series. Here the state at $\nu = 4/11$ is perhaps the most interesting one\cite{pan}, and it is important to test our proposed wave function numerically. To do this efficiently one must work on a finite geometry, which in practice means a sphere or a torus. Recently, we have shown how to formulate the wave functions \pref{psin} on a torus\cite{torus}, and preliminary numerics on the $\nu = 4/11$ state is encouraging. 

\begin{acknowledgments}
We thank Juha Suorsa for interesting discussions, and Eddy Ardonne for  useful comments on the manuscript. 
This work was supported by the Swedish Research Council and by NordForsk.
\end{acknowledgments}

\end{document}